# ASSOCIATIVE DETACHMENT (AD) PATHS FOR H AND CN⁻ IN THE GAS-PHASE: ASTROPHYSICAL IMPLICATIONS.

Stanka V. Jerosimić[1] , Franco A. Gianturco[2] , Roland Wester[2]

[1] Faculty of Physical Chemistry, University of Belgrade, Studentski trg 12-16, PAC 105305, 11158 Belgrade, Serbia

[2] Institut für Ionenphysik und Angewandte Physik, Universität Innsbruck, Technikerstraße 25, 6020 Innsbruck, Austria

**Abstract**

The direct dynamical paths leading to Associative Detachment (AD) in the gas-phase, and specifically in the low-temperature regions of the Dark Molecular Clouds (DMC) in the ISM, or in cold trap laboratory experiments, are investigated with quantum chemical methods by using a high-level multi-reference Configuration Interaction (CI) approach that employs single and double excitations plus Davidson perturbative correction [MRSDCI(Q)] and the d-aug-cc-pV5Z basis set. The potential energy curves for H + CN⁻ are constructed for different directions of the H partner approaching the CN⁻ anion within the framework of the Born-Oppenheimer approximation. The present calculations found that the AD energetics at low temperature becomes favorable only along a selected range of approaching directions, thus showing that there is a preferred path of forming HCN at low temperatures, while that of forming its HNC isomer is found to be energetically forbidden. Given the existence in the ISM of different HCN/HNC ratios in different environments, we discuss the implications of our findings for selective formation of either isomer in the low-temperature conditions of the Molecular Cloud Cores.

*corresponding author: francesco.gianturco@uibk.ac.at

1. INTRODUCTION

The possible mechanisms of interstellar formation of hydrogen cyanide and hydrogen isocyanide, HCN/HNC, have been limited early on to the electron-initiated chemistry of $H_2CN^+$ [1], whereby the latter molecule was shown to play a role in both dense and diffuse interstellar clouds. Observationally, the HCN/HNC isotopes have been confirmed in a variety of astrophysical environments: in diffuse and translucent interstellar clouds [2], in dense interstellar clouds [3], in star-forming regions [4], protoplanetary disks [5], external Galaxies [6], and in comets [7]. An interesting feature from such a wide range of observation has been the marked variations of the isomer ratios in different astronomical environments. In the dense interstellar clouds, where the gas is shielded from external UV starlight, the abundance ratio was found to be close to 1.0 [8], while in regions exposed to UV radiation the HCN isotopologue was found to be more abundant than HNC by a factor of 5, both in diffuse clouds [2] and in photon-dominated regions (PDRs) [9]. Such marked differences must be linked to elementary mechanisms involving formation/destruction along selective paths which would be distinguishing among the two isomeric species: we shall argue below that the associative detachment (AD) path could be one of such selective reactions.

The two molecules in question are also strongly polar systems, with a dipole moment ($\mu$) of 2.984 D for HCN [10], and computed to be 2.866 D for HNC [11]. It will play an important role for the present analysis since it turns out that HCN has an electron affinity (EA) which is negative for the traditional valence-bound (VB) anion state [12], while its permanent dipole moment is large enough to allow for the existence of a weakly-bound dipole-bound (DB) located outside the molecule and spatially very diffuse [13]. Its predicted and measured EA value is around 1.56 meV for the molecule in its $j=0$ rotational state, while it would dissociate fully for $j=3$: a rotationally "hot" HCN molecule would not even support a DB state [13]. For the HNC DB state, the binding energy is somewhat larger, i.e. around 6.70 meV [11], but still fairly small.

Another observed molecule, rather ubiquitous in the ISM and in dark and diffuse molecular clouds, is the $CN^-$ molecular anion [14], which contrary to the two isomeric forms of hydrogen cyanide, has a very robust VB anionic state: EA of 3.862 eV [15]. Its bond length is only 0.01 Å longer than its neutral counterpart, and its ground state is a closed-shell compact system of $X\,^1\Sigma^+$ symmetry. We can therefore see that this diatomic anion and the corresponding 3-atomic anions discussed earlier are very different molecules with respect to electron-initiated processes. They can however be linked through a reaction involving the ubiquitous hydrogen atoms found in the interstellar environments:

$$\text{CN}^- (X\,^1\Sigma^+) + \text{H}\,(^2S) \leftrightarrow \text{HCN}\,(X\,^1\Sigma^+) + e^- \quad (1)$$

The above reaction has been extensively studied computationally because of the challenge provided by the instability of the hydrogen cyanide to electron attachment processes [11,16]. It is essentially a recombination reaction that is exothermic to forming the neutral system and therefore markedly favoured at the low temperatures of many of the ISM environments. Qualitatively we can say that the compact, valence excess electron on CN$^-$ is being replaced by a hydrogen atom, thereby producing in the three-atom system a diffused excess electron that is weakly bound to the products and which can be easily detached, thus leaving behind the energetically preferred configuration of a neutral molecule plus an outgoing continuous electron.

The mechanism corresponds to an Associative Detachment (AD) path to products [17–19]. Once measured in a plasma discharge, the experimental value of the AD coefficient of this reaction was reported to be $8\cdot 10^{-10}$ cm$^3$s$^{-1}$ [20]. The same reaction is also known to occur for producing the isomeric variant:

$$\text{NC}^-(X\,^1\Sigma^+) + \text{H}\,(^2S) \leftrightarrow \text{HNC}\,(X\,^1\Sigma^+) + e^-. \quad (2)$$

It is also known that the AD reactions involving Hydrogen atoms and its symmetric (D$_2$) and asymmetric (HD) products

$$\text{H}^- + \text{H} \leftrightarrow \text{H}_2 + e^- \quad (3)$$

do not have energy barriers to reaching the AD final products and therefore their rates steadily increase down to very low temperature values, in line with molecular hydrogen (H$_2$) production in dark and diffuse molecular clouds [21].

The questions we wish to answer aim at linking the behavior of reactions (1) and (2) in the low-T regimes to the consequences they might have about the survival, or the destruction, of the CN⁻ anions observed under interstellar environments and the appearance of the neutral products of the isotopologues HCN/HNC. In particular, we shall carry out accurate quantum structure calculations, described in detail in the following Section, which will allow us to argue, for some specific aspects of its electronic energy changes, the importance of these reactions on the selective abundances of the concerned partners.

These questions can be summarized as follows:

(*i*) the analyzed AD path leads to a region of anion instability within the coordinate space of vibrationally "cold" HCN and HNC. Is this region directly accessible without barriers or are energy barriers present at some geometries, thereby preventing neutral product formation at the low-T regimes of the Dark Clouds?

(*ii*) Is there any evidence of a selective behavior of the two isomeric products, HCN and HNC, thereby suggesting different efficiencies for their formation via the AD mechanism?

(*iii*) Can one envisage the existence of an intermediate complex for (HCN⁻)* and (HNC⁻)* on their way to the electron detachment metastability regions? Can we also explain the electronic mechanism of the formation of a weakly bound excess electron that will auto-detach at the nuclear geometries close to the equilibrium structure of HCN/HNC ?

The present paper is structured as follows: Section 2 will provide the details of the quantum calculations and describe the quality of the employed *ab initio* methods. The following Section 3 will present our calculations for the different angular "cuts" of the 3-atom potential energy surface (PES). Section 4 will finally present our conclusions and outline their effect for the ISM abundances of the title molecules.

## 2. DETAILS OF QUANTUM CALCULATIONS.

The Geometries of the HCN/HNC system, and of CN⁻ interacting with the approaching H atom, are described by using Jacobi coordinates to present the paths of the incoming atom H towards CN⁻ from different angles (see Fig. 1). In that figure the $R$ coordinate is the distance of H to the center of mass of the $^{12}$C and $^{14}$N, and in our calculations goes from 4.5 Å to about 1 Å, depending on the angle $\theta$. All calculations were carried out in the $C_s$ point group and additionally using the $C_{2v}$ point group at linearity. In our choice of coordinates, the linear approach to the nitrogen atom corresponds to $\theta = 0°$, while the corresponding linear approach to the carbon atom is at $\theta = 180°$.

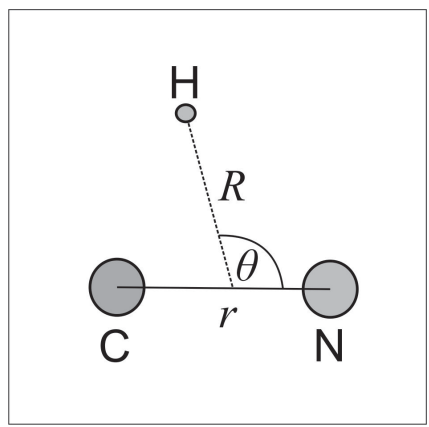

Fig. 1 Jacobi coordinates of triatomic molecular complex of the present study. The hydrogen atom moves along the $R$ coordinate taken from the center of mass of the CN system, while the $\theta$ coordinate varies from the linear HCN ($\theta=180°$) to the linear HNC ($\theta=0°$) complexes.

The present Potential Energy Surface (PES) describes the energetics for the interaction of the CN⁻ anion in its ground state ($X\ ^1\Sigma^+$) with the atom H, leading to a neutral HCN plus the outgoing continuous electron. Before we describe the method for obtaining the PES used in this work, let us mention previous results that are of importance for guiding us in choosing the level of theory employed in our work. We have already mentioned that HCN⁻ is a dipole-bound (DB) anion, which does not form a valence-bound (VB) negative ion: HCN bounds an excess electron chiefly by electrostatic interaction and very weakly bound anions are extremely often found in nature.

Petersen and Gutowski [22] calculated the electron binding energies of HCN and HNC with coupled-cluster methods, reported very small binding energy and found unusually large electron correlation energy contributions. It was reported elsewhere that coupling the electron motion to rotational motion of the nuclei is of critical importance to the HCN⁻ spectrum [13,23], and that the Born–Oppenheimer (BO) treatment can be inadequate in predicting bound state spectra of this weakly-bound anionic species. This concern, however, turns out to be more significant when specifically searching for bound rotationally excited states of the anionic complex, a task outside the scopes of the present study.

We are not intending to investigate the bound states of HCN⁻ and of its excited states at nuclear geometries where the neutral HCN has the pronounced minimum on its ground state potential . Rather we are analyzing this system from a different point of view from that already presented in Chourou et al. [12], where the dissociative electron attachment to stable HCN or HNC was analyzed directly for the formed complexes. Instead, we are modeling H atoms approaching stable CN⁻ species in the ISM environments at geometries initially away from the bound HCN/HNC potential well (i.e. outside of the auto-detaching region). Therefore, we can confidently apply the multi-reference quantum structural calculations with high level of dynamical correlation and with extra-diffuse basis functions for describing the initial AD paths of main interest in our study.

Electronic energies for the lowest-lying adiabatic states of H−CN and H−CN⁻ systems were calculated by using the state-averaged (SA) full-valence complete active space self-consistent field (CASSCF) reference function for the subsequent internally contracted multi-reference configuration interaction calculations with single and double excitations (MRCISD) [24,25], corrected with perturbative Davidson correction [MRCISD(+Q)]. In the state-averaged procedure both the lowest neutral and the lowest anion states were included. One-electron space was described by using the doubly-augmented polarized valence correlation consistent 5-zeta basis set (d-aug-cc-pV5Z) [26] with 431 contractions that include two diffuse functions of each symmetry, specifically invented for describing electrostatic properties, van der Waals binding forces, dispersion forces, etc. All geometry optimizations and energy calculations were performed using the same method (sometimes also using a smaller d-aug-cc-pVTZ basis set). In the present study, the full valence space consists of 10 electrons in 9 valence orbitals; in the MRCI calculations the core electrons were not optimized. Calculations were carried out using MOLPRO 2012.1 software package.[27,28]

Potential energy curves for the lowest electronic adiabatic states of the neutral HCN/HNC and HCN$^-$/HNC$^-$ were calculated using the C−N distance kept fixed at the equilibrium geometry of the CN$^-$ anion. By using our computational method, we obtained for the latter the value of 1.1817 Å (the equilibrium geometry of the $X\,^1\Sigma^+$ state).

We further analyzed the anionic H-CN$^-$ system using information obtained from electronic structural calculations, to find the possible van der Waals complex H⋯CN$^-$, and the molecular features of it that might affect the mechanism for the AD reaction in the low-T range of the ISM environment. The curves obtained at $\theta=180°$ were also compared to the single-reference RCCSD(T) curve using the same basis set. At distances smaller than about $R=3$ Å we obtain essentially the same shape of the anionic curve, although the difference of the asymptotic values from this CC method (3.24 eV) deviate both from our calculations of the electron affinity (EA) of CN done at the MRCI(Q) level (3.854 eV), and from the existing experimental value (3.862 eV)[15].

## 3. COMPUTATIONAL RESULTS.
### 3.1. Different 'cuts' of the full PES: implications on the AD .

The ground state of the neutral HCN is $X\,^1\Sigma^+$ state ($1\,^1A'$ in the $C$s group), which originates from the dominant configuration $[\text{core}](3\sigma)^2\,(4\sigma)^2\,(5\sigma)^2\,(1\pi)^4$, with the equilibrium geometry of r(CN)= 1.1571 Å, r(CH)= 1.0615 Å. The second electronic state is $^1\Sigma^-$ (vertically reached), in fact a bent $^1A''$ state which lies about 6.5 eV above the ground state.[29] Geometry optimization of the lowest-lying doublet state of HCN$^-$ ($^2\Sigma^+$ state, or $1\,^2A'$) in the vicinity of the equilibrium geometry of HCN (its electronic configuration is $[\text{core}](3\sigma)^2\,(4\sigma)^2\,(5\sigma)^2\,(1\pi)^4\,(6\sigma)^1$, where 6σ is a very diffuse orbital) predicts nearly the same geometry as that obtained for the neutral HCN. However, the computed EA of HCN at its equilibrium geometry (and generally at distances $R$<2 Å) is usually found to have a negative value, thus indicating metastability of the triatomic anion, apart from the formation of the weakly bound, Dipole-bound (DB) state. It therefore follows that at such geometries of the 3-atom system the weakly bound electron can undergo autodetachment from the anionic complex, as the latter becomes the less stable nuclear configuration, thereby decaying into the ground state of the neutral HCN plus a continuum electron.

Hence, the quantum chemistry calculations for the potential energy curves (PEC) at a fixed angle for the anion become insufficient at distances smaller than the crossing point ($R$< $R_D$): here $R_D$ denotes the value of the $R$ coordinate where, physically speaking, the electron detaches from the anion leaving behind the neutral species. The calculations for the full system, therefore, are not yielding any real-valued energy quantities: the physical states of the anionic complex reached at such geometries become metastable and acquire imaginary components related to the lifetime of the (n+1)-electron state involved. They will be presented as dots in our figures below.

Concerning the neutral species, its ground electronic state is well separated from the other states of the 3-atom system. On the other hand, the three lowest states of the A' symmetry of the anion exhibit several avoided crossings at the SA-CASSCF level under $R$=2 Å. The present $^2\Sigma^+$ state of the anion is the lowest-lying adiabatic state.

In Figs 2-5 the lowest lying electronic states of this species are presented for different values of $\theta$ and for $R$ values up to an outer distance of 4.5 Å: the latter distance was considered sufficient to effectively represent the interacting species as "asymptotically" apart from each other.

In Figure 2 the radial potentials of neutral and anionic states are presented for linear HCN/HCN¯ system ($\theta$=180º), linear HNC/HNC¯ ($\theta$=0º), and for angles $\theta$ = 150º, 120º, 90º, and 60º. The $^2$A' ($^2\Sigma^+$ at linearity) state of HCN¯ system is the electronic state that correlates with the ground H($^2$S) and the ground CN¯ ($^1\Sigma^+$) asymptotic states; therefore, the asymptote in our figures (as $R$→∞) corresponds to H+CN¯, while the asymptote for the neutral $^1\Sigma^+$ corresponds to the ground H+CN.

The energy difference between these two asymptotes obviously equals the energy difference between the ground state of CN¯ and that of the CN. The neutral HCN/HNC asymptote lies 3.854 eV above the HCN¯/HNC¯ asymptote, very close to the experimentally obtained difference between CN¯ and CN radical of 3.862 eV [15]. The crossing point between the neutral $1^1A'$ and anion $1^2A'$ states changes with the angle $\theta$. The energy of the ground states of H + CN¯ lies above the energy of the bound HCN/HNC which indicates that the AD reaction is an exothermic process, as already pointed out in our Introduction Section.

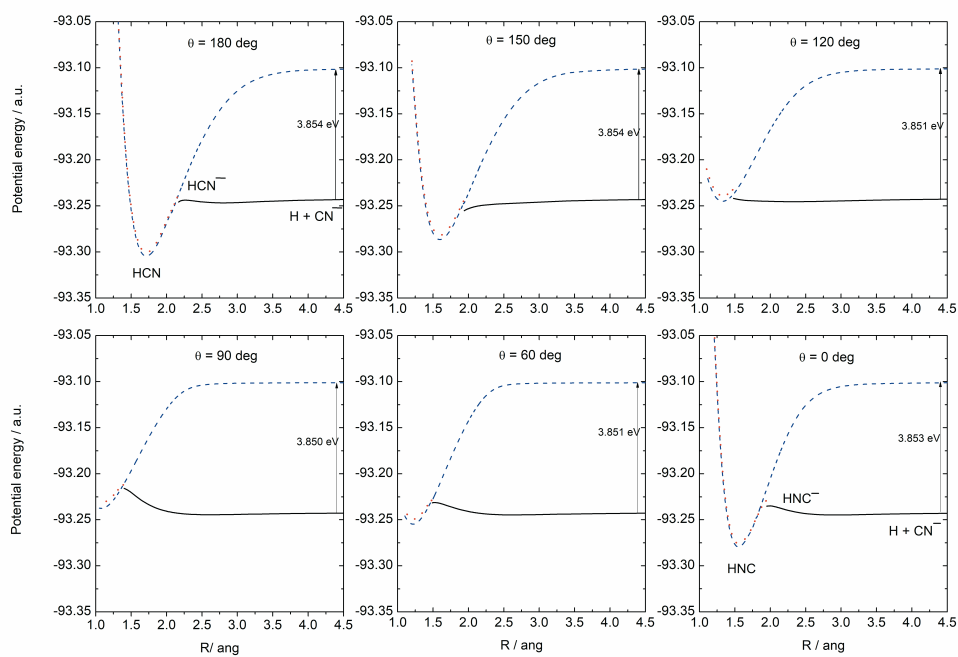

Figure 2. Computed total energies of the lowest-lying adiabatic states of neutral HCN/HNC and of the anion HCN¯/HNC¯ species, for different values of $\theta$, along the $R$ coordinate, at a fixed CN distance of 1.1817 Å. The values were calculated at the SA-CASSCF-MRCISD(Q)/d-aug-cc-pV5Z level. The anionic, stretched H---CN¯ system is shown by solid lines, while the neutral H-CN curve is given by dashed lines. The anionic system in the physical region of geometries smaller than the point D, where the stable neutral has a lower energy than its metastable anion, is given by dots.

We shall now focus on enlarged views of the specific segments of the PECs presented in Figure 2, namely, the stretched H---CN¯ and the possible presence of energy barriers as partners approach the anionic equilibrium geometry , where the AD reaction occurs.

For better clarity, we present in Figure 3 two specific branches of the PECs in units of wavenumbers as an astronomically more familiar energy unit. The angles 180º and 165º are shown as examples. We shall help the discussion by marking important points along the curves from A to D.

As further discussed below, van der Waals complexes for the H···CN¯ are formed near the $R$ = 2.85 Å and over a specific range of angular approaches. We should note that, at least in principle, such structures could become bound states, through stimulated Radiative Association (RA) mechanism in the collision-less regime of the Interstellar regions. We shall further discuss below the question of their long-term stability and of their possible importance for the present reaction. One should note, however, that the spontaneous/stimulated RA rates for systems with small dimensionality, like the one we are considering, usually yield at low temperatures rather small values (e.g. see ref. [30]) and would therefore play a marginal role for the present AD reaction.

For the specific examples of Figure 3, however, we see that the reference asymptotic energy (point A) is well above the well regions in both cases, thus indicating that the increase of relative kinetic energies, which occurs between partners as they approach the geometries of the location B through an attractive potential, can overcome the formation of bound states and also the energy of the barriers at the points C. It then follows that approaches along these PECs would bring the partners without hindrance into the region of the AD process. We shall further discuss this point below.

For the collinear approach shown in Figure 3, the basis set superposition error (BSSE) pertaining to our basis set was computed using the counterpoise method [31,32]. For that linear complex at $R$=2.8 Å (the point B), we obtained a value of −0.13 kJ mol$^{-1}$ (−16 K), which was 0.7% of the total value of the binding energy. At the point D in Figure 3, where the fragments H and CN⁻ are closer to each other, the calculated value was of −0.32 kJ mol$^{-1}$ (−38 K).

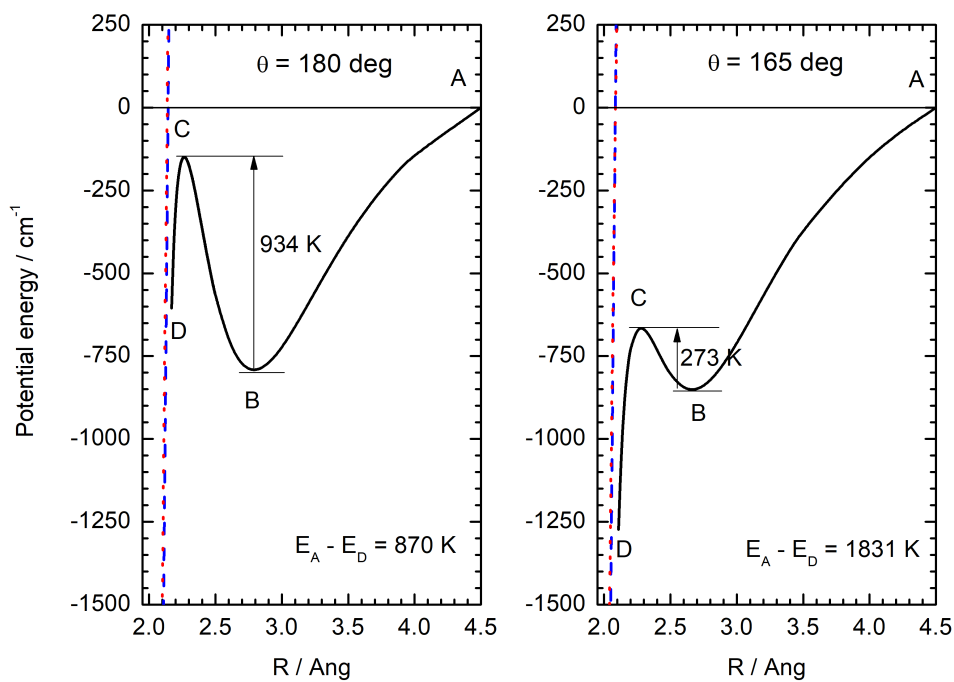

Figure 3. PECs for $\theta$=180º and 165º, along the $R$ coordinate, at fixed CN distance of 1.1817 Å, calculated at the SA-CASSCF-MRCISD(Q)/d-aug-cc-pV5Z level. The anionic, stretched H---CN⁻ system is shown by solid lines, while the onset of the neutral HCN curve is given by the dashed vertical lines of its attractive branch. The anionic complex at the metastable geometries is given by dots. See main text for further details.

In Figures 4 and 5 we further show the enlarged views of the PECs for additional values of $\theta$. In Figure 4, the PECs over that group of smaller angular values (from 160º to 130º), have no intermediate bound complex and show no barrier along the path to the metastable region while in Figure 5, and for each of the PECs (from 120º to 0º), an energy barrier clearly appears on the path to the inner region of metastability. Just as importantly, the data of that Figure also show that the asymptotic energy values (the locations A) are lower than the points D where one enters the metastability region of the AD path. The significance of such findings will be discussed below, while the energies at the important locations for the PECs at different Jacobi angles are summarized in Table 1 to provide more quantitative information. We report there the results of our total energy values in hartree, as is conventional in Quantum Chemistry studies, while the relative differences are given in kelvin for better understanding their possible kinetic significance within the low-T regimes of the ISM. It is worth mentioning here that for some of the angles the corresponding PECs support stationary points for which the contributions of zero-point vibrational energies (ZPVE) can be evaluated, as further discussed below.

Table 1. Energy values at selected locations along the anionic PECs, depicted in Figures 3-5, and at different angles: point A defines the 'asymptotic' energy ($CN^-$ + H), point B identifies the complex energy minimum, C indicates the location of the barrier, D locates the closest distance for which the AD region is reached. We present the total energies in a.u. (hartree) and the relative energies in Kelvin. The zero of the energy corresponds to the selected "asymptotic" value at point A.

| angle | $R_A$ / Å<br>E(A)/a.u. | $R_B$ / Å<br>E(B)/a.u.<br>E(B)-E(A)/ K | $R_C$ / Å<br>E(C)/a.u.<br>E(C)-E(A)/ K | $R_D$ / Å<br>E(D)/a.u.<br>E(D)-E(A)/ K |
|---|---|---|---|---|
| 180 | 4.5<br>−93.24309258<br>0 | 2.8<br>−93.24672721<br>−1148 | 2.26<br>−93.24377076<br>−214 | 2.17<br>−93.24584711<br>−870 |
| 165 | 4.5<br>−93.24305749<br>0 | 2.7<br>−93.2469482<br>−1229 | 2.3<br>−93.24608232<br>−955 | 2.11<br>−93.24885564<br>−1831 |
| 160 | 4.5<br>−93.24303215<br>0 | 2.5<br>−93.2472179<br>−1322 | 2.35<br>−93.24717199<br>−1307 | 2.06<br>−93.25119921<br>−2579 |
| 150 | 4.5<br>−93.24296825<br>0 | − | − | 1.94<br>−93.25517841<br>−3856 |
| 135 | 4.5<br>−93.24286074<br>0 | − | − | 1.69<br>−93.25613878<br>−4193 |
| 120 | 4.5<br>−93.24276701<br>0 | 2.35<br>−93.24540144<br>−832 | − | 1.48<br>−93.24178031<br>+312 |

| 105 | 4.5 −93.242703730 | 2.6 −93.24456944 −589 | – | 1.43 −93.22451954 +5742 |
|---|---|---|---|---|
| 90 | 4.5 −93.242674470 | 2.6 −93.24438633 −541 | – | 1.39 −93.21201111 +9683 |
| 75 | 4.5 −93.242676080 | 2.6 −93.24443679 −556 | 1.44 −93.22011354 +7125 | 1.41 −93.22052109 +6996 |
| 60 | 4.5 −93.242702350 | 2.6 −93.24456439 −588 | 1.55 −93.23122712 +3624 | 1.49 −93.23156856 +3516 |
| 45 | 4.5 −93.242745130 | 2.7 −93.24465014 −602 | 1.72 −93.23833497 +1393 | 1.62 −93.23979414 +932 |
| 30 | 4.5 −93.242793620 | 2.8 −93.24465087 −586 | 1.88 −93.23836173 +1399 | 1.74 −93.24149717 +409 |
| 15 | 4.5 −93.242832870 | 2.9 −93.24462606 −566 | 1.95 −93.23598178 +2163 | 1.78 −93.24561843 −880 |
| 0 | 4.5 −93.242848060 | 2.9 −93.24461573 −558 | 2.00 −93.23477212 +2550 | 1.95 −93.2350288 +2469 |

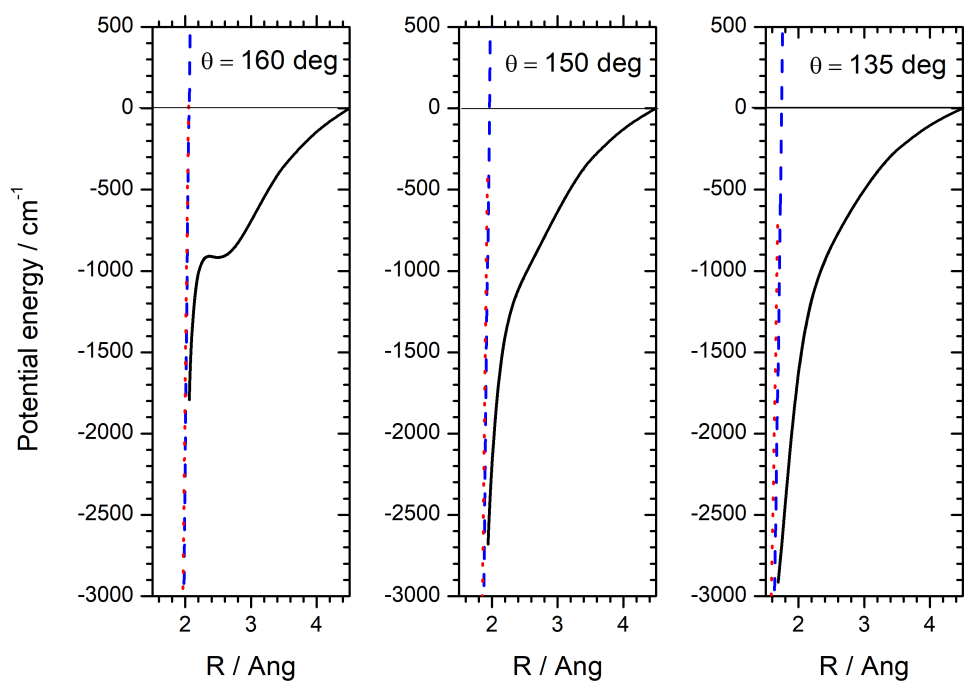

Figure 4. PECs for θ=160º, 150º, and 135º along the R coordinate, at the fixed CN distance of 1.1817 Å, calculated at the SA-CASSCF-MRCISD(Q)/d-aug-cc-pV5Z level. The anionic, stretched H---CN⁻ system is shown by solid lines, while the attractive branch of the neutral HCN

curve is given by the dashed lines. The autodetaching region is given by dots. See main text for further details.

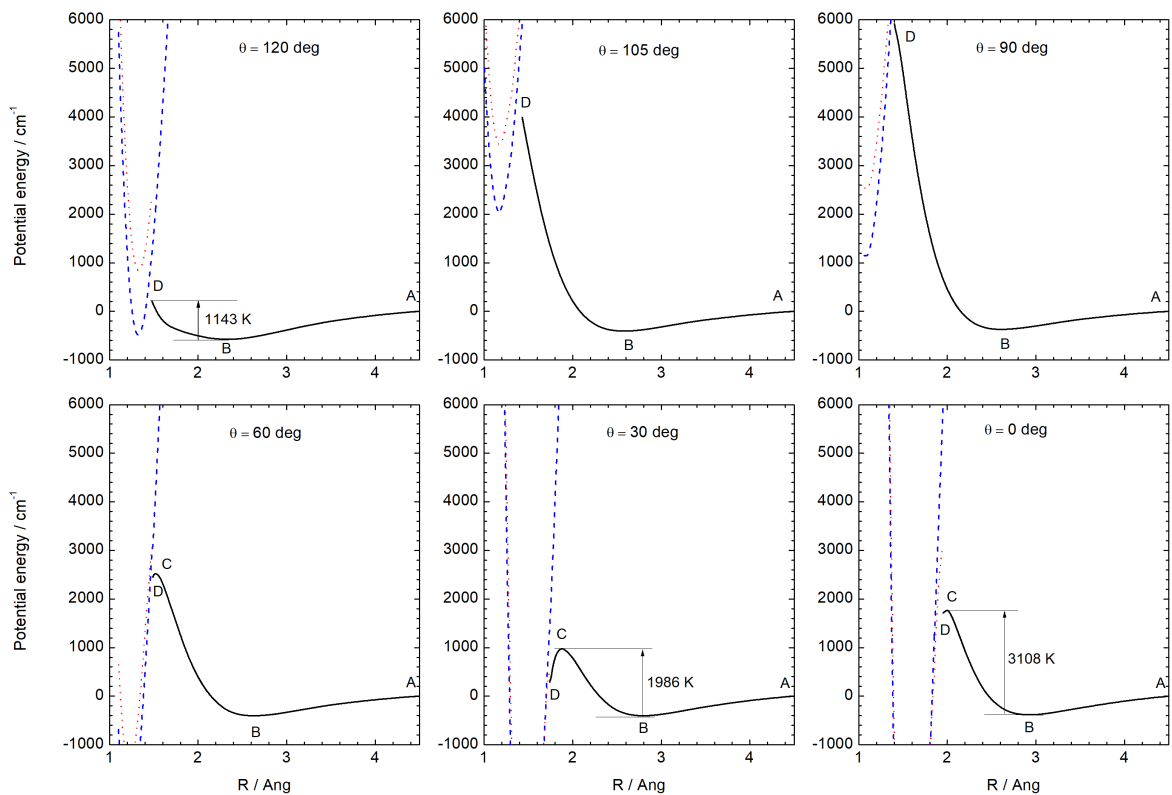

Figure 5. The same type of computed data as those in Figure 4, but here reported for the smaller range of angles. Top panel row: θ=120º, 105º, and 90º; bottom panel row: θ=60º, 30º, and 0º. See main text for further details.

Thermodynamically, the reaction of AD formation of HCN/HNC from the interstellar interaction of H and CN⁻ species can take place whenever the point D is lower in energy than point A, the one that labels the selected asymptotic energy in each curve: both are given in our Table 1. Additionally, we see that the possible barrier, labeled by its height at the point C, can prevent the partners to reach the metastability region from the minimum energy of that complex and whenever the latter is formed along the selected angular orientation of the PECs. Thus, low-temperature encounters in the cold regions of the ISM will not be likely to bring the reacting partners to further reach those geometries of the complex which would lead to the excess electron's detachment into the continuum.

We see in Table 1 that $E_A > E_D$ from $\theta =180º$ up to about 120º: From the smaller range of angles, where $E_A < E_D$, the opposite occurs and the AD reaction becomes endothermic down to the angular values where the H approaches from the N-side of the ionic partner. Furthermore, the barrier that appears when the incoming H atom is moving from the points labeled B to C is as high as 934 K at 180º and up to 273 K at 165º (see Fig. 3), while it becomes smaller when the system is bent, and from 160º (up to about 130º) the present calculations predict that no barrier should occur to the the metastability region of partners' geometries (Figure 4).However, as already mentioned, the large exothermicity at the largest angles will actually overcome barriers like those of Figure 3 since the initially low- kinetic energy partners will gain velocity down the attractive potential PECs existing at those angles. The AD reactions can therefore occur.

For angles smaller than 120º , however, we have seen that the reaction becomes endothermic and the existing energy barriers vary from 1,143 K at 120º up to 10,223 K at 90º. Hence, for slow colliding partners approaching each other along such trajectories the extra kinetic energy which might be gained in going from A to B would not be sufficient to overcome the barriers located at points C or at the D "gates" to the metastability geometries: to reach that region, where the AD reaction could occur, becomes therefore forbidden.

In the regions where H is approaching $CN^-$ on the N side ($\theta$ from 90º to 0º), the barriers from the minimum B to the point C become even larger: from 7,681 K at 75º, to 3,108 K at 0º. At $\theta=15º$ we further see that the point D is below the energy value at the geometry of point A, i.e. the reaction becomes again exothermic . However, the barrier from point B to the top of point C is still large (+2,730 K) and not likely to be overcome by the small relative kinetic energy increase between the cold-environment approaching partners. Hence, the AD reaction is once more prevented from occurring at the low temperatures of interest.

Thus, the colliding partners which approach each other along relative geometries within a sizeable angular cone around the nearly collinear HCN can enter without energy barriers the AD reaction region. The smaller angles of approach, on the other hand, pertain to the endothermic regimes: closer to the T-shape configuration the approaching paths exhibit now increasingly larger energy barriers and have asymptotic energy values lower than the locations of points D . The low-T regimes will therefore prevent the complex to enter the regions where the excess electron could move to a (DB) state of the complex from which it would undergo the autodetachment process.

The calculations therefore suggest that the AD path of $CN^-$ destruction leading to HNC formation is energetically forbidden at low temperatures and will not happen easily in the environment of the Molecular Clouds where $CN^-$ has been observed.

### 3.2. Features of the Van der Waals complex between H and $CN^-$.

An interesting aspect of our study is the possibility to analyze the electronic structures of the stretched anionic $H-CN^-$ complex at different points on the PECs presented in Figures 3 and 4, in order to gain some additional information on the molecular mechanism of the reaction for which we assume small relative kinetic energy between partners.

A possible $H\cdots CN^-$ van der Waals complex would in fact lie on the same potential energy surface. One may then surmise that its possible formation could prevent the stretched system to undergo further deformations into the AD reaction region because of possibly forming bound states via a stimulated Radiative Association (RA) mechanism, a stabilizing process which could be present in the collision-less ISM environment. We shall show below, however, that such bound states may decay by angular deformations and therefore not likely to play an important role.

Our present calculations show that, at large separations of H and $CN^-$, the extra electron occupies the degenerate valence $6a'$ and $1a"$ orbitals of the $CN^-$ anion ($\pi^4$ config.) and that the HOMO (the highest occupied molecular orbital) of the whole system, the $7a'$ orbital, consists of one electron centered on the hydrogen, 1s (H). As H is approaching the CN⁻, the character of the $6a'$ and $1a''$ orbitals do not change significantly, but the character of the $7a'$ orbital changes and our calculations show that, just before entering the metastable region of geometries, it contains AO contributions from all three atoms; the largest coefficient (at θ=180º) pertains to a diffuse 7s AO of C, then $6p_z$(C), 1s(H), 6s(H), 8s(N), etc. Therefore, the shape of this orbital changes from being centered on the H atom (in its 1s orbital), to becoming distributed on all three atoms and having a very diffuse molecular shape within the complex. Skurski et al. [11] had also found that, along such a path, the extra electron smoothly changes its spatial features from being a compact valence electron to becoming a more diffuse DB electron, and that the H atom then experiences stronger binding on the C than on the N side of CN⁻ anion when one compares the two isomeric structures. They had not found a van der Waals minimum and hence concluded that there is no barrier separating such a complex from the HCN⁻ as H approaches CN⁻ along a path from the C side of CN⁻ partner for bent geometries: the same effects can be seen from our Figure 4.

At linearity, however, those authors have found a stationary point (at $R$ = 2.76 Å) with a binding energy for the linear complex (relative to the asymptote H + CN⁻) of 3.0 kcal mol⁻¹ (1,510 K). The latter complex is separated by a barrier of 1.2 kcal mol⁻¹ (604 K) from the metastable region where electron autodetachment can occur. It also turns out that this linear complex is unstable with respect to bending deformations. Our calculations indeed confirm the existence of this stationary point for the linear H-C-N geometry and with $R$= 2.857 Å, $r$(CN)= 1.1846 Å, with the value of dipole moment of 1.5307 D, and with harmonic frequencies given as follows:

1) $v_1$ = 282 cm⁻¹ which corresponds to the stretching H−CN⁻, where H is oscillating along the molecular axis, with the calculated intensity of 209 km mol⁻¹,

2) $v_2$ = $i$63 cm⁻¹ (imaginary frequency of the H−CN⁻ bending mode, where mostly H is moving in the direction normal to molecular axis at linearity, and is of low intensity),

3) $v_3$ = 2053 cm⁻¹ (the C-N stretching mode, with low intensity of 2 km mol⁻¹).

This complex structure is, therefore, a transition state along the bending coordinate and it is also the minimum structure regarding the stretching modes. We can estimate the zero-point vibrational energy to be of 13.96 kJ mol$^{-1}$ (1,679 K), while omitting the bending mode. Our value of the barrier to the metastability region is found to be 934 K (Figure 3, Table 1), and the calculated value for the binding energy taking into account the counterpoise correction [32] is −18.7 kJ mol$^{-1}$ (2,252 K) at the geometry with $R$= 2.8 Å, $r$(CN)=1.1817 Å and $\theta$=180º.

The calculated value for the harmonic frequency of the CN$^-$ anion is of 2060.65 cm$^{-1}$, to be compared with an earlier theoretical value of 2057 cm$^{-1}$ [34] and with the experimental value of 2068.6 cm$^{-1}$ [34]), with the zero-point vibrational energy (ZPVE) of 12.33 kJ mol$^{-1}$ (1,482 K). We can also see that along the $R$ coordinate towards stable HCN, the ZPVE value of the anionic complex becomes larger and, at the point B, it amounts to about 1,679 K. However, as the nuclei approach each other more closely, e.g. at the geometries corresponding to points C or D, the vibrational energy at zero K ( hence assuming very small relative kinetic energy values) cannot be sufficiently large to overcome the electronic energy barrier that exists at linearity and therefore it would prevent the system from leaving the complex.

In order to better analyze the possible AD path near linearity, and at the kinetic conditions indicated before, we additionally performed geometry optimizations starting from different points on the PES and along the angular cone within which AD reaction can occur on the C-side (from $\theta$=180 to 135º). Optimizations from θ=135º, 140º, 150º and $R$=3 Å lead to the region where the AD reaction can occur without local minima and with no barrier being detected as H approaches the CN$^-$. Figure 6 shows PEC shapes around the angular region where a stationary point was found: the lowest-lying points of each angular 'cut' are the D-points already defined earlier. We see in the Figure that, starting from the linear complex unstable under the bending motion, the interaction energy is going down slowly and, after 160º, it is sloping downward until neutral HCN is formed and the extra electron is ejected into the continuum. The conclusion from the above structural considerations is that for gas-phase partners approaching each other in the low-temperature regions along linear configurations there is no significant energy barrier toward the formation of a stable neutral HCN, because the unstable complex H-CN⁻ can find its way to the autodetaching region without encountering any barrier.

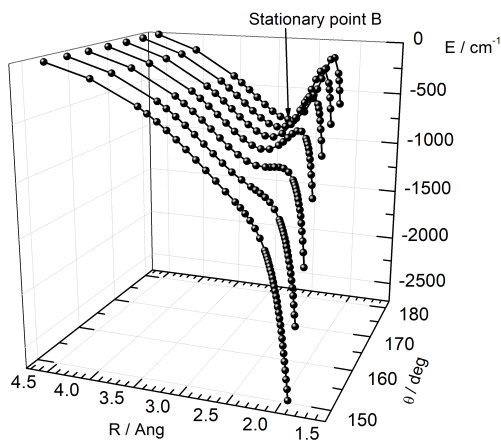

Figure 6. Potential energy surface cuts at different angles of the anionic complex H-CN⁻ before entering the metastable region (on the right of the plots). The angular values go from $\theta = 180°$ to $150°$. The lowest-energy points for each Jacobi angle are the D-points defined in the main text and in Table 1.

**3.3. The role of charge distributions and dipole moments along the AD paths.**

In order to further discuss in a simple, qualitative way how the bound charges (i.e. the physical electrons within the complex) are moving along the potential curves when nearing the linear configurations of the three atoms, we present the Mulliken population analysis. The latter has been calculated from the density matrix which was in turn obtained from MRCI charge distributions of the lowest anionic electronic state.

We show our results for different geometries, i.e. for various $R$ values at two different angular values in the two panels of Figure 7: The left-side panel is for $\theta = 150°$ and the right-side panel for $\theta = 180°$. We expect that such a simplified analysis realistically traces the changes of the partial atomic charges, i.e. the changes of the total number of electrons which can be allocated at each atom or the charge-per-atom when the total charge of complex is obviously −1.

From the data in that Figure we see that the complex behaves very similarly for both geometries. At the nearly-asymptotic distance of H+CN$^-$ ($R$=4.5 Å), H, C and N have the (excess) charges of 0, −0.22 and −0.78, respectively. At the smaller $R$ distances, the excess charge is also largely located on the N atom while both C and H have fairly small portions of the excess electron charge distribution. However, the process of having the CN$^-$ partner interacting with the incoming H atom moves the smaller portion of excess electron from N to C and from C to H. In other words, we see that at the D points defined earlier the excess charges are now as follows: −0.11, −0.13, −0.76 for H, C, and N, respectively, at 150º, and −0.02, −0.36, −0.62 at 180º. After reaching the metastability geometry at R<R$_D$, the charge values become very large and negative on the H atom and large and negative on the N atom. As we pointed out earlier, the physical reality is that these structures are not stably bound since at such geometries the neutral HCN becomes the more stable molecule: our simple indicators, therefore, are telling us that a very diffuse electron is about to be formed on our metastable complex and that it will be now moving away from the whole molecule, starting chiefly from the H-side of the complex, to eventually turn into a detached, continuum electron and leaving a stable neutral product from the AD reaction. It is therefore only the use of scattering calculations that could provide a physically more realistic picture for the electron detachment into the continuum, although our quantum chemical calculations are already pointing to the correct molecular description of the AD mechanism.

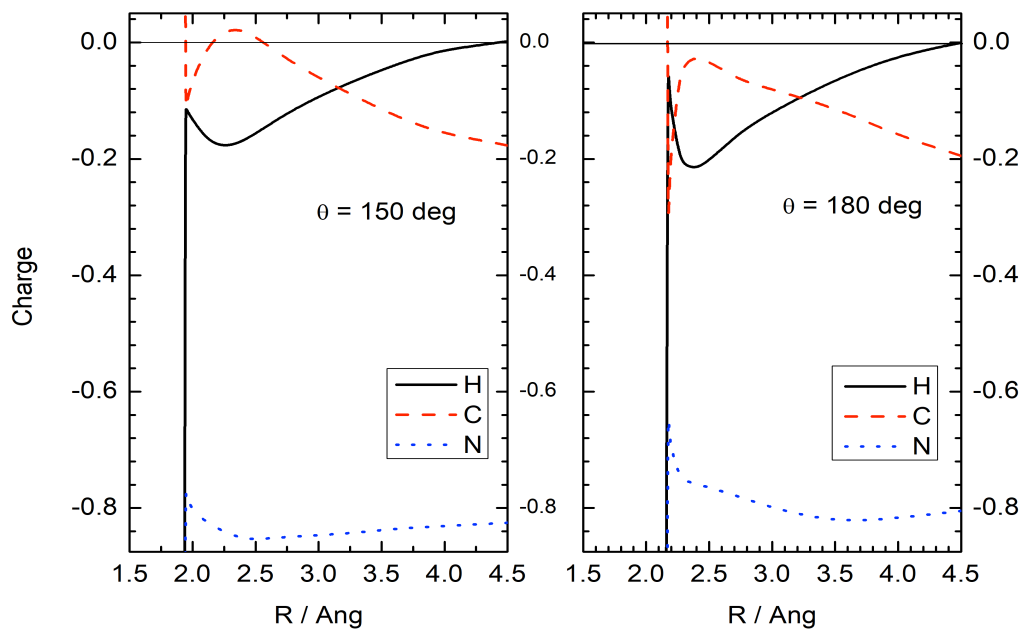

Figure 7. Computed Mulliken population analysis from the density matrix from MRCI charge distributions of the lowest anionic electronic state and for different radial geometries at two fixed θ angles. Left panel: θ =150º, right panel: θ= 180. See main text for further details.

In Figure 8 we additionally present, besides the changes of charge distributions in its top panel, the changes of the dipole moment magnitudes in the anionic complex and for the neutral H-CN. They are shown in the middle panel of that Figure, while on the bottom panel we report again the PEC for $\theta$=150º already discussed earlier. Note that the direction of the vector of the electric dipole moment in the anionic complex and in its neutral counterpart have opposite behavior: in the neutral molecule it points from N→C→H and therefore the overall negative pole is on the N-side of the molecule, while in the anionic complex the vector points along the H→C→N direction so that the overall negative pole is now on the H and C-side of the molecule. Figure 8 also shows that the dipole moment of the anionic complex increases in value as the H atom is approaching the CN⁻ partner and that, once the radial region of metastability is reached (at $R$=1.94 Å) it very rapidly shoots up to a very large unphysical value of more than 15 D. The latter is again an indication of the fact that the HOMO electron is becoming spatially very diffuse and that it is entering the complex region where autodetachment is occurring and the neutral molecule is preferentially formed. Thus, an excessively large value for its dipole moment is indicating that the negative end of the charge distributions is moving away from the molecule, thereby producing such large values from the existing MO coefficients.

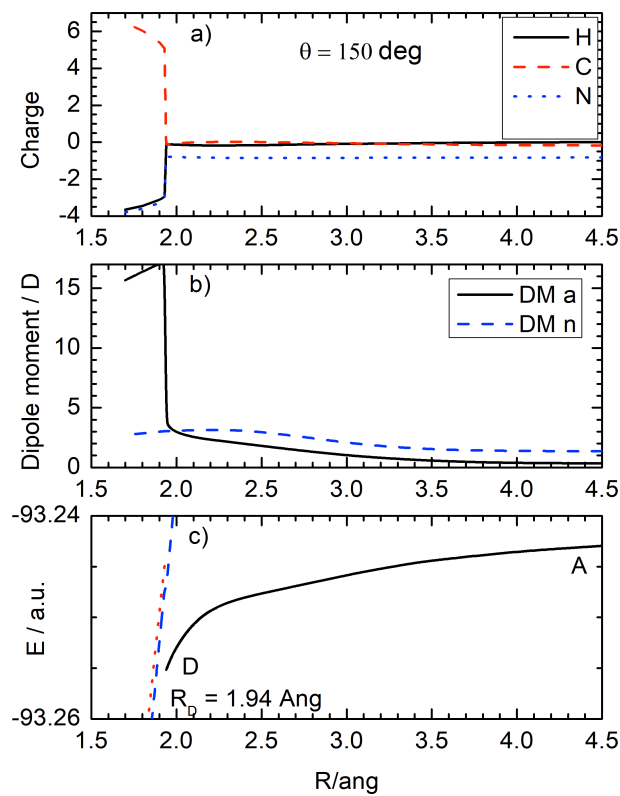

Figure 8. Mulliken charges, electrical dipole moment norms of the anionic and neutral complexes, and potential energy curve, at $\theta=150°$ reported as a function of the radial relative distance for a fixed angular value shown in the top panel. See main text for further details.

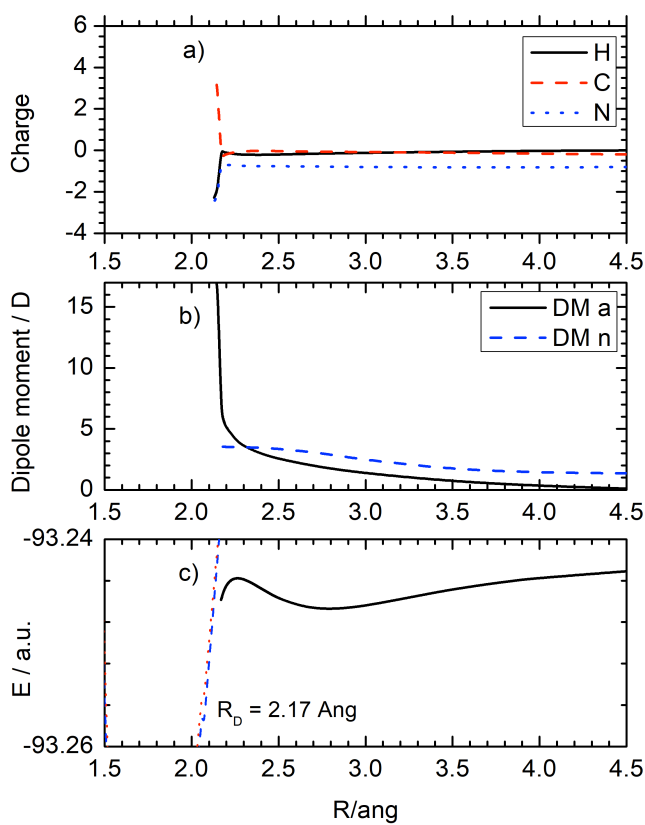

Figure 9. Mulliken charges, electrical dipole moment norms of the anionic and neutral complexes (they have opposite directions), and potential energy curve, at $\theta=180º$, reported as a function of the radial relative distance for that fixed angular value. See main text for further details.

The data reported by Figure 9 are presenting results for the full linear complex configurations. We see essentially the same behavior of all the quantities shown and the role of the quantum chemical observables in trying to provide indicators for the electron-detachment process.

## 4. PRESENT CONCLUSIONS

In this work we analyze in detail the molecular mechanism of reaction (1) in the gas phase and under relative kinematic conditions which would be similar to those in the cold regions of the ISM where the stable anion CN⁻ has been observed. The computational methods have been those of *ab initio* Quantum Chemistry, using a high level of accuracy for correlation and polarization effects.

This study looks at the molecular mechanisms of the AD reaction, causing destruction of the $CN^-$ and formation of neutral HCN/HNC molecules. The calculations show that the interacting partners reach a distance of approach where the anionic complex becomes metastable and a neutral triatomic molecule plus a continuum electron is the energetically favoured species. Furthermore, calculations clearly indicate that there are marked stereodynamical effects, since different directions of relative approach of the partners can give rise to energy barriers before the anionic complex reaches the geometries where the autodetachment occurs, this being the case when the AD reaction also becomes endothermic. Thus, we have found a specific angular region ( between 120° and 0°) for the approaching H and $CN^-$ which strongly prevents, in the low-temperature, the interacting partners from reaching the AD region. This occurs when the intermediate complex would preferentially lead to one of the isomeric species: the HNC neutral molecule. As a consequence of it, the two constitutional isomer forms of the HCN would have different evolutionary histories in the ISM environments since, while HCN can be formed even at low temperatures via the AD reaction mechanism, the HNC formation would be prevented by an energy barrier. Such difference of behaviour provide an additional reason as to why the relative ratios of HCN/HNC can greatly vary in different regions of the ISM, as discussed in detail in the Introduction Section. In other words, our calculations indicate that the molecular properties of the two isomers are favouring formation of one of them, but not of the other. Hence, the present $CN^-$ destruction path via AD would lead more efficiently, at low temperatures, to forming the HCN than the HNC molecule.

The final products are also expected to be formed in their vibrational ground states, while their rotational temperatures may be "hotter" and therefore provide $j>0$ final states of the intermediate three-atom complexes, thereby favoring electron detachment in the metastable region [13].

By analyzing different molecular properties of the two complexes, as they approach the AD region, we were also able to show that the polar nature of the molecules and the EA values of the neutrals involved, play a significant role in the AD mechanism.

DB configurations of the excess electron can be formed and the weakly-bound anionic complex can undergo loss of the excess electron. We further found that this weakly bound electron has been localized chiefly to the H-end of the triatomic complex before entering the autodetaching region and has become a very diffuse metastable electron: this is the quantum chemical picture of an intermediate state of this bound electron on its way to the molecular electronic continuum of the residual HCN/HNC.

## ACKNOWLEDGEMENTS

F.A.G. and R.W. thank the FWF, Austrian Science Fund (FWF) for supporting the present research through the Project P27047-N20. This work was also supported by the Project No. ON-172040 awarded to S.J. by the Ministry of Education, Science and Technological Development of the Republic of Serbia. All authors further acknowledge the financial support of the COST Action COST-STSM-CM1401-34125 for the award of a STSM to SJ for a visit to the University of Innsbruck, where the present work was initiated.

**GRAPHICAL ABSTRACT**

The Associative Detachment reaction between H and CN$^-$ at low temperature becomes possible only along a selected range of approaching directions, thus showing that there is a preferential possibility at low temperatures of forming HCN in comparison with forming CNH.

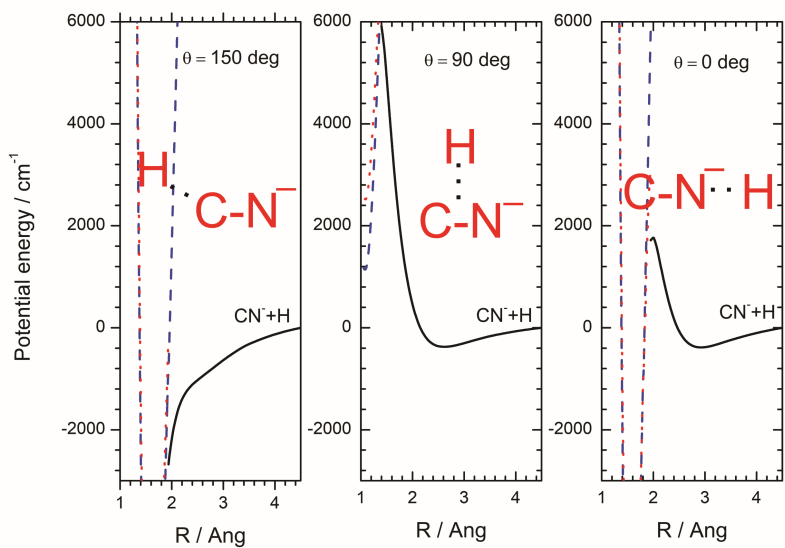